# THE NEUTRON-TAGGING FACILITY AT LUND UNIVERSITY

F. MESSI, H. Perrey, K. Fissum, D.D. DiJulio, E. Karnickis, V. Maulerova, N. Mauritzson, E. Rofors,
Division of Nuclear Physics, Lund University and
European Spallation Source ERIC
Lund, Sweden
Email: francesco.messi@nuclear.lu.se

A. Huusko, T. Ilves, A. Jalgén, S. Koufigar, H. Söderhielm, D. Söderström
Division of Nuclear Physics, Lund University
Lund, Sweden

R. Hall-Wilton
European Spallation Source ERIC
Lund and
Mid-Sweden University
Sundsvall, Sweden

P. Bentley, C.P. Cooper-Jensen, J. Freita-Ramos, F. Issa, K. Kanaki, A. Khaplanov, G. Mauri, F. Piscitelli, I. Stefanescu
European Spallation Source ERIC
Lund, Sweden

J. Scherzinger
Department of Physics, University of Pisa
Pisa, Italy

R. Al Jebali
European Spallation Source ERIC
Lund, Sweden and
School of Physics and Astronomy, Glasgow University,
Glasgow, UK

J.R.M. Annand, L. Boyd
School of Physics and Astronomy, Glasgow University,
Glasgow, UK

M. Akkawi, W. Pei
University of Toronto,
Toronto, Canada

**Abstract**

Over the last decades, the field of thermal neutron detection has overwhelmingly employed He-3-based technologies. The He-3 crisis together with the forthcoming establishment of the European Spallation Source have necessitated the development of new technologies for neutron detection. Today, several promising He-3-free candidates are under detailed study and need to be validated. This validation process is in general long and expensive. The study of detector prototypes using neutron-emitting radioactive sources is a cost-effective solution, especially for preliminary investigations. That said, neutron-emitting sources have the general disadvantage of broad, structured, emitted-neutron energy ranges. Further, the emitted neutrons often compete with unwanted backgrounds of gamma-rays, alpha-particles, and fission-fragments. By blending experimental infrastructure such as shielding to provide particle beams with neutron-detection techniques such as tagging, disadvantages may be converted into advantages. In particular, a technique known as tagging involves exploiting the mixed-field generally associated with a neutron-emitting source to determine neutron time-of-flight and thus energy on an event-by-event basis. This allows for the definition of low-cost, precision neutron beams. The Source-Testing Facility, located at Lund University in Sweden and operated by the SONNIG Group of the Division of Nuclear Physics, was developed for just such low-cost studies. Precision tagged-neutron beams derived from radioactive sources are available around-the-clock for advanced detector diagnostic studies. Neutron measurements performed at the Source Testing Facility are thus cost-effective and have a very low barrier for entry. In this paper, we present an overview of the project.

# 1. INTRODUCTION

Neutrons of all energies are important probes of matter. The precise detection of neutrons emerging from a sample under study is crucial to the quality of the resulting experimental data. Until recently, He-3-based technologies were essentially the only method used for neutron detection. The recent He-3-crisis [1,2] and the proposal of the European Spallation Source (ESS) [3, 4] have led to an aggressive search for alternative technologies [5]. Together, the prohibitive cost of He-3 and the design goals for the new facility with its extremely high flux of neutrons call for completely new concepts for detectors [6] and shielding [7]. Strong candidates for new detector technologies exist, but few of these have been characterized properly. Most are still in their developmental infancy and need to be precisely validated. In general, the validation of a new detector technology is a two-step process: first, wide-ranging irradiations are performed using neutron-emitting radioactive sources, where the cost per neutron is low; and second, promising technologies are then precisely irradiated at neutron-beam facilities, where the cost per neutron is substantially higher. We note that the cost per neutron at a neutron-beam facility can be so high that it may be prohibitively expensive. A facility based upon neutron-emitting radioactive sources is thus a cost-effective solution to this problem with a relatively low entry threshold. Once the sources and necessary infrastructure are in place, "natural" neutrons are available around the clock. Thus, the overhead for initially benchmarking new technology will not be dominated by beam-time associated costs. Further, by instrumenting a source facility with well understood shielding and equipment to take advantage of nuclear-physics knowledge associated with the particular radioactive decay in question, low-cost polychromatic beams of neutrons may be created.

At the Division of Nuclear Physics at Lund University [8], the internationally accessible Source-Testing Facility (STF) [9] facility has been constructed to provide precision beams of neutrons from radioactive sources. Thus, a cost-effective solution for performing advanced detector diagnostics already exists and is in fact routinely employed by its users. The STF has been instrumented to provide all the tools necessary for the initial characterization of newly developed detectors and shielding materials to its users.

# 2. THE SOURCE TESTING FACILITY

Constructed at the Division of Nuclear Physics at Lund University in collaboration with the Detector Group [10] at ESS, the STF is operated by the SONNIG group [11]. It is a fully equipped user facility. The STF boasts a complete range of neutron and gamma-ray sources for the characterizations of detectors and is also equipped with several detectors, detector-associated infrastructure and IT, as well as the electronic components essential to the needs of a user-focused laboratory. As there are no reactors or accelerators involved, the STF is available continuously for prototype development and commissioning. Figure 1 shows an overview of the facility.

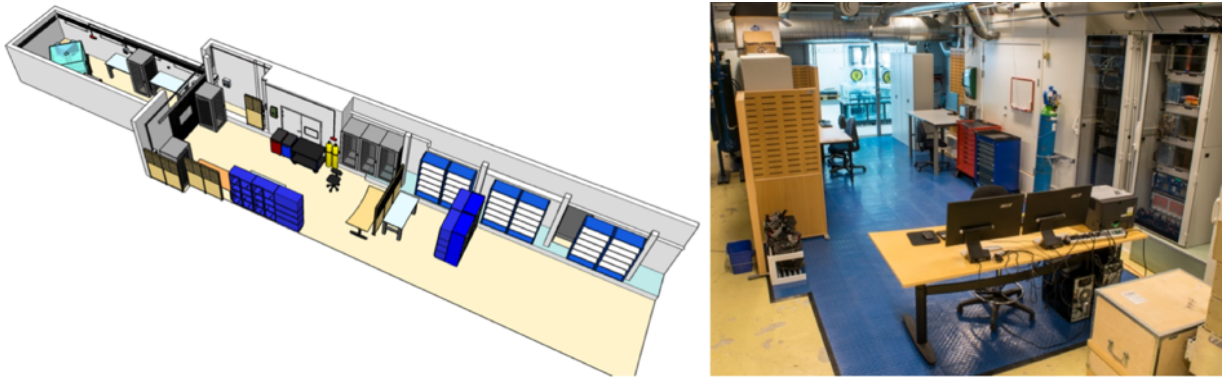

*Figure 1: The STF. On the left, a 3D-rendering and on the right, a photo.*

The key infrastructure available at the STF includes:

- **Sources**: several mixed-field Actinide-Beryllium sources, a fission neutron source as well as gamma-ray sources. In particular, $^{241}$Am-Be, $^{238}$Pu-Be and $^{252}$Cf (thin window) sources are available as well as $^{57}$Co, $^{60}$Co, $^{137}$Cs, $^{22}$Na and $^{133}$Ba sources for gamma-ray radiation.

- **Detectors:** - a full set of detectors, both commercially provided and in-house developed, for background monitoring, gamma-ray detection and fast- and thermal-neutron detection. The detector pool of the STF includes plastic and liquid scintillators, gas detectors (He-3 tube, He-4 cells), solid-boron detectors, inorganic crystal scintillators (such as 1.5" CeBr$_3$ and LaBr$_3$ as well as a 12" NaI) and boron-straw tubes, to name a few. Example of detectors available at the STF are shown in Figure 2.

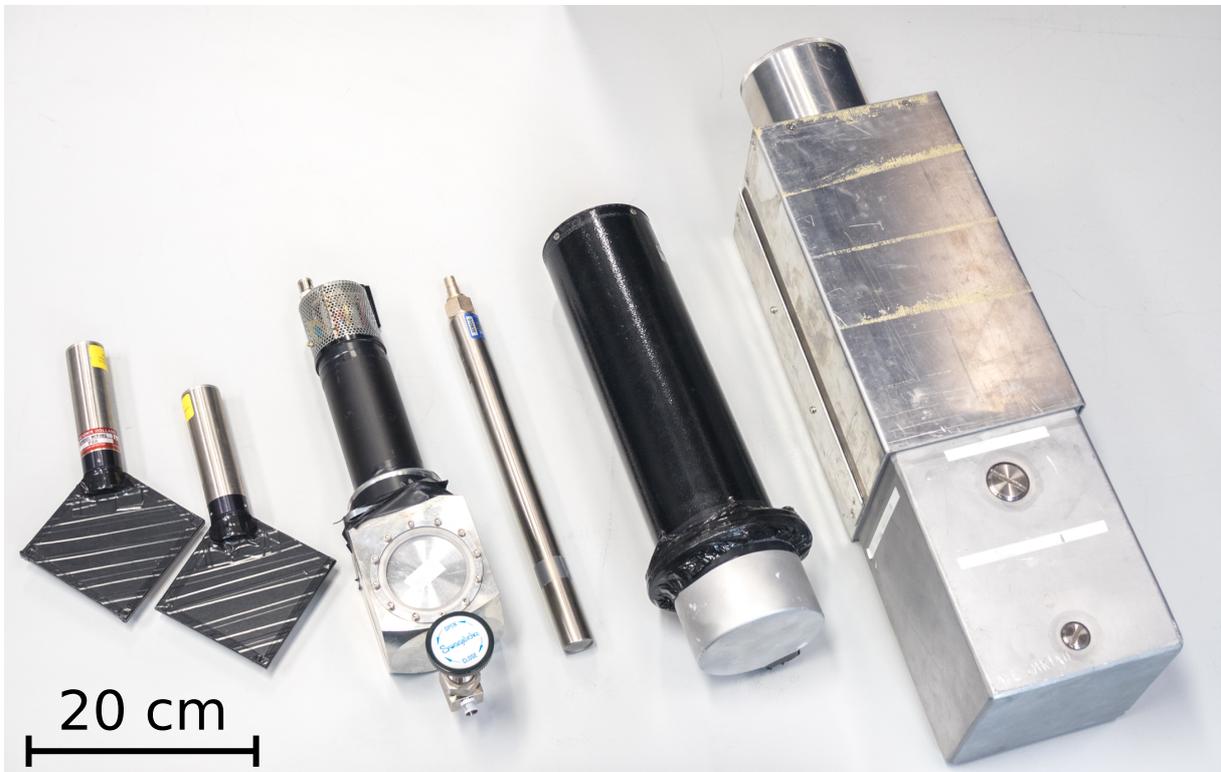

*Figure 2: Example of detectors available at the STF. From the left, two plastic scintillator paddles read out by PMTs, one He-4 cell (operated at 5 atm) read out by a 2" PMT, a He-3 tube (calibrated efficiency of 96.1% at 2.5Å) and two NE213 liquid scintillator (0.43 and 4.5 l) read out by a 3" PMT.*

- **The Aquarium**: a custom-designed shielding apparatus for neutron sources, delivering if desired beams of "tagged" neutrons[1] (see Figure 3). The Aquarium consist of a 3-section cube of Plexiglas (~1.4 m side), filled with about 2650 l of high-purity water. It is designed to host a neutron source in its centre, together with four gamma-ray sensitive detectors. The dose-rate on the external surface of the cube is < 0.5 μSv/h when an industry standard 18.5 GBq $^{241}$Am-Be source is encapsulated. Four horizontal cylindrical apertures of ~17 cm in diameter act as "beam guides", one perpendicular to each of the four vertical faces of the cube, providing four uniform but combined beams of gamma-rays and neutrons from the source.

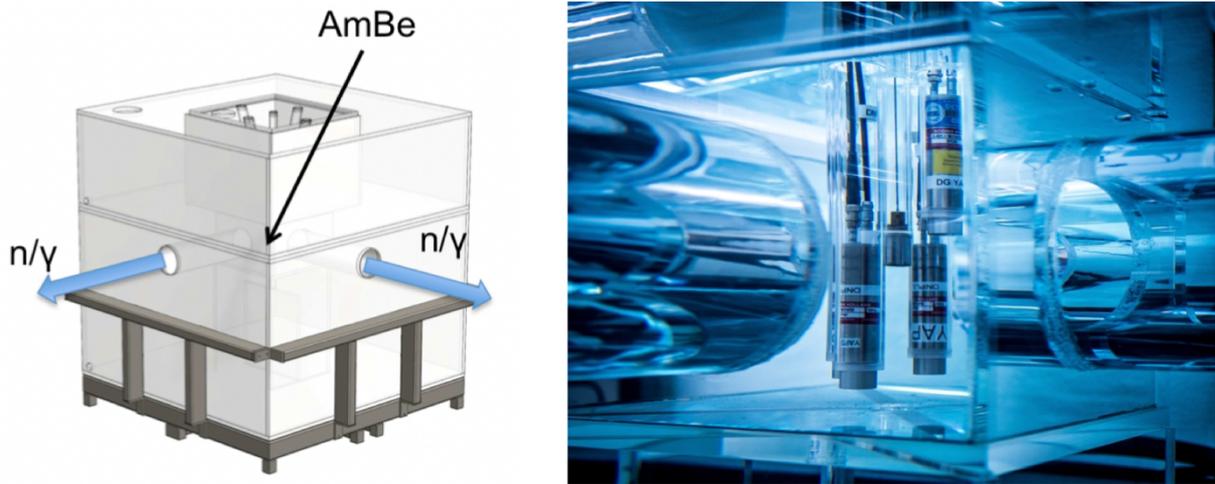

*Figure 3: The Aquarium. On the left, a CAD drawing and on the right, a photo of the inner chamber of the Aquarium. The neutron source surrounded by four gamma-ray detectors as well as two of the four beam ports may be seen.*

- **A black-box**: a light-tight enclosure for testing light-sensitive detectors such as open photo-multiplier tubes (PMTs) (see Figure 4). The enclosure will eventually contain an optical table with a 1 m$^2$ work surface. One end of the enclosure will house servo stages which can either carry a calibrated laser emitter or radioactive sources. These servo stages will allow for the mapping of the topological response of areal detectors such as multi-anode PMTs.

---

[1] The neutron-tagging technique is described in detail in Sec. 3.

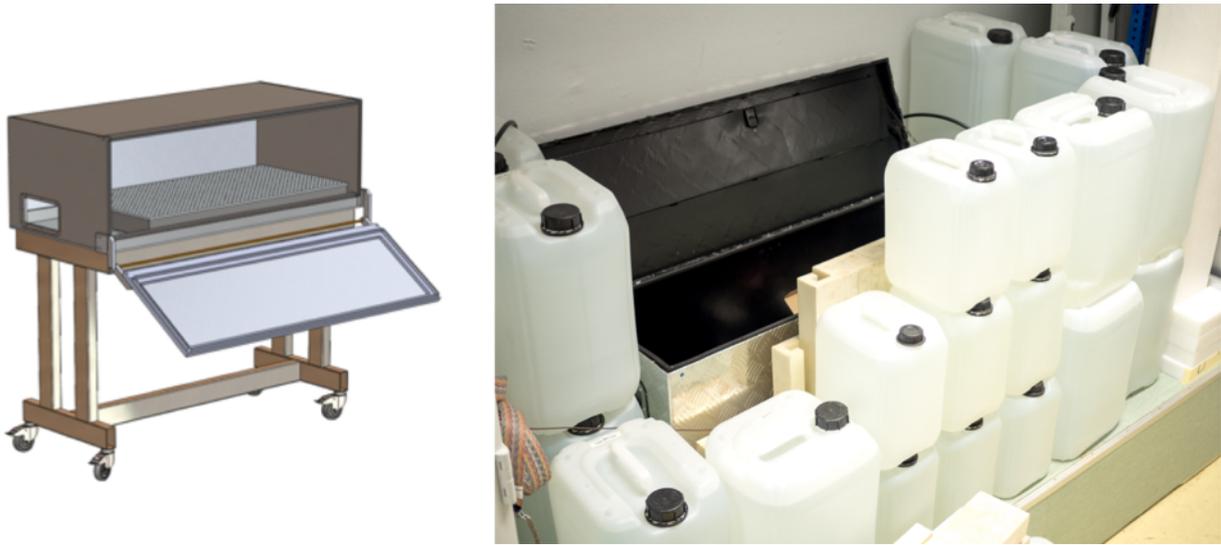

*Figure 4: The black-box at the STF. On the left, a CAD drawing of the apparatus under construction and on the right, a photo of the existing prototype. The light-tight box may be surrounded by modular shielding.*

- **Electronics and computers**: the STF is equipped with a comprehensive set of electronics modules and computers. The facility is designed to be modular and flexible, so that users can bring their own equipment or use the available infrastructure in any combination. This includes analog NIM, CAMAC and VME modules (discriminators, QDCs, TDCs, visual scalers, etc...) as well as more modern digitizers (see Figure 5). Several computers are also available to be connected to the experimental setup under consideration to acquire data from the detectors.

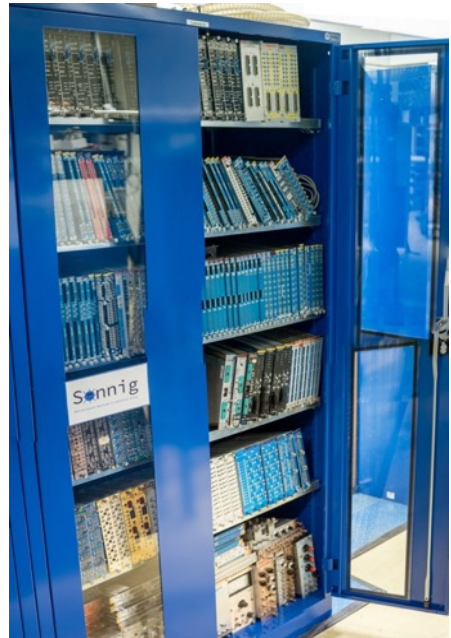

*Figure 5: A subset of the electronics modules available at the STF.*

- **Shielding materials:** such as plastic or lead bricks, as well as borated-Al plates and various geometries of borated-plastic material. These may be used to optimise experimental setups.

- **Acquisition and analysis software** are both available, in particular a pair of ROOT-based DAQs [12]. The DAQs may be employed to collect data and to provide a first analysis (see Figure 6). Commercial software is also available for various MCA and Digitizer modules.

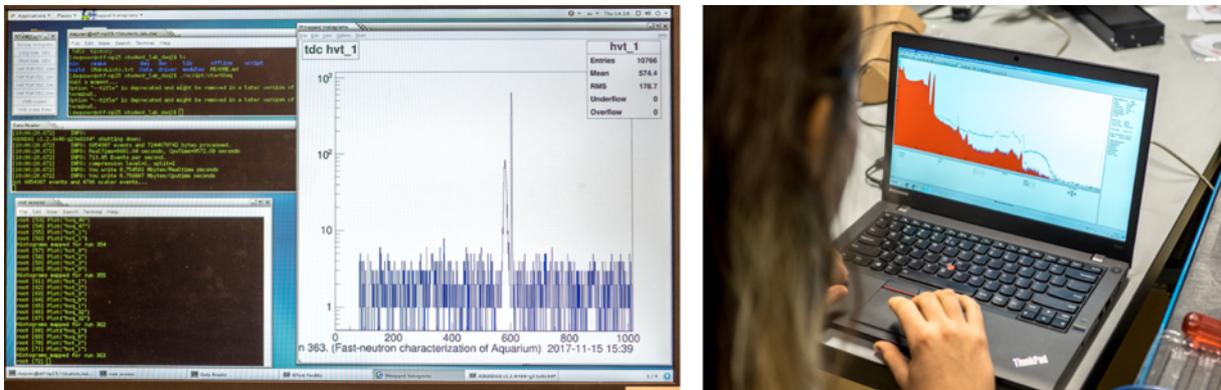

*Figure 6: DAQs running at the STF. On the left, a screen shot of a ROOT-based DAQ and on the right, a student using the MCA software.*

- **Simulation tools**, based on GEANT4 [13] and included in the simulation framework of the ESS [14, 15, 16], have been developed to characterize the Aquarium and the sources. They are intended to be used to facilitate understanding of features within data that would be difficult to study in real life.

- **SONNIG expertise.** The group members are available for consulting and support. They are highly experienced with the set-up of the experiment and/or the optimization of the DAQs and with how to use the analysis software. Experience with the setup is gladly shared and help with optimizing acquisition software as well as data analysis can be provided.

## 3. TAGGING NEUTRONS

Employing radioactive sources for detector characterizations can be advantageous to the user. For example, once the setup is optimised and the acquisition of data is started, in contrast to accelerator-based measurements, no further assistance is needed and no night shifts are required. A disadvantage of employing radioactive sources is that the emitted neutrons have a wide range of energies that are not uniform. The ISO 8529-2 recommended neutron-energy spectrum from $^{241}$Am-Be is shown in Figure 7. As can be seen, the emitted neutrons are definitely not mono-energetic. Moreover, any neutron-emitting source likely emits a mixed field of gamma-rays, alpha-particles and neutrons.

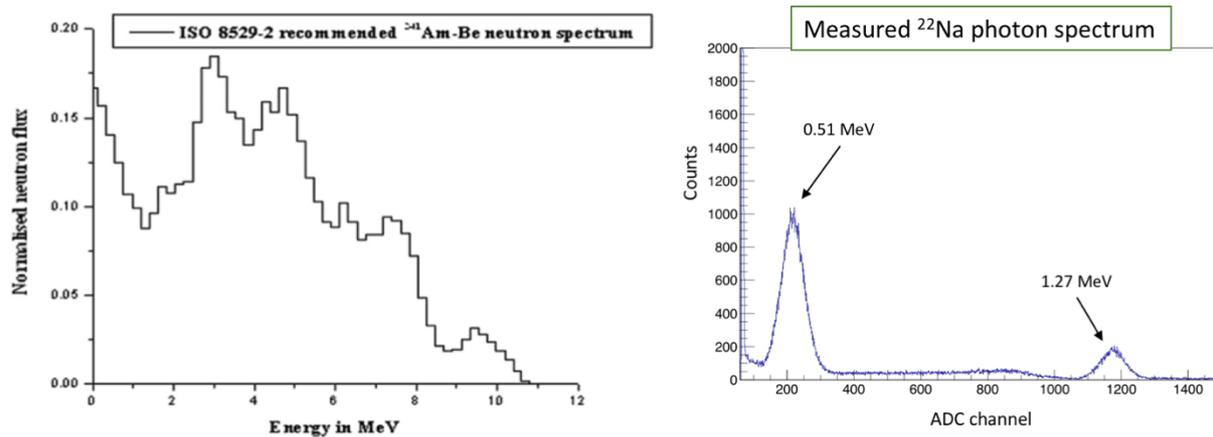

*Figure 7: Spectra from sources. On the left, the ISO 8529-2 recommended neutron spectrum for a $^{241}$Am-Be source and on the right, a measured gamma-ray-spectrum for a $^{22}$Na source.*

Further, neutron sources generally radiate mixed field isotropically. Due to the mixed field, the wide energy spectrum of the released neutrons, and the randomness of the underlying decay processes, direct-exposure irradiations offer a less controlled environment compared to a reactor beam line. Clearly, such a beam line may be carefully tuned to provide a continuous mono-energetic beam of neutrons. However, by precisely measuring the radiation field on an event-by-event basis, one can reconstruct the properties of each individual neutron and thereby "tag" the neutrons. The process involves determining the time-of-flight (ToF) and thus the energy of each detected neutron [17, 18].

In a source, such Am-Be, fast neutrons are emitted via the reaction
$$\alpha + {}^9Be \rightarrow {}^{12}C + n$$
The recoiling $^{12}$C is left in its first exited state about 55% of the time and the emitted neutron is accompanied by the prompt emission of a 4.44 MeV gamma-ray from the instantaneous de-excitation of the $^{12}$C to its ground state. If both the neutron and the gamma-ray are detected, the ToF and thus kinetic energy of the neutron may be determined on an event-by-event basis (see Figure 8).

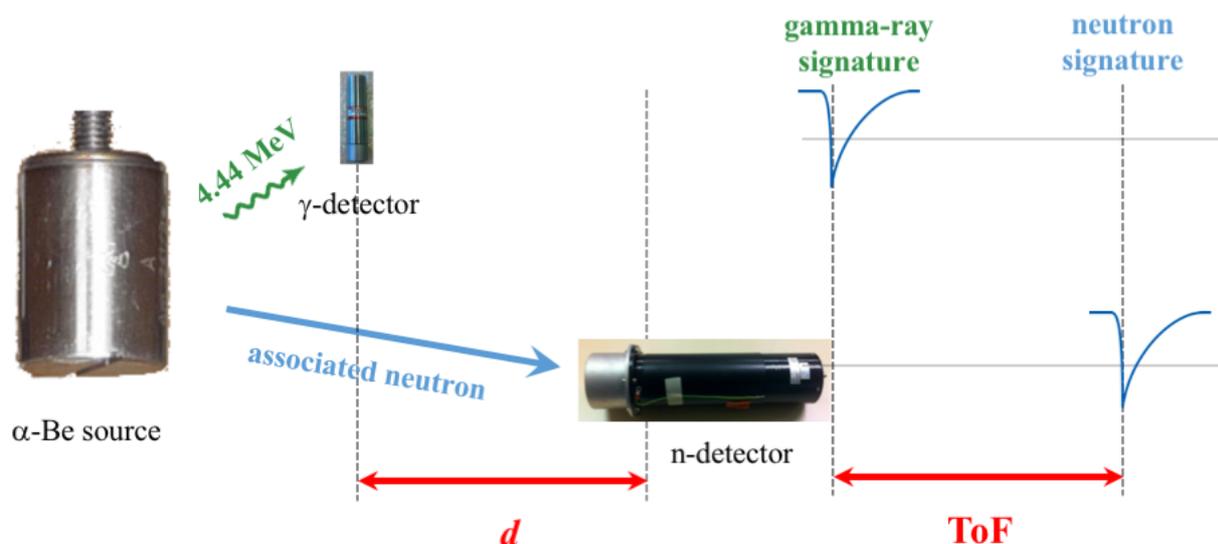

*Figure 8: Neutron-Tagging technique. A 4.44 MeV gamma-ray is measured in conjunction with the associated emitted neutron. From the time difference between the detection of the two particles, the ToF of the neutron can be determined and thus its energy may be calculated.*

In our case, the relative timing between the detection of the gamma-ray by an Yttrium Aluminum Perovskite (YAP) inorganic crystal-scintillator detector and the detection of the neutron by an organic liquid-scintillator detector is measured, resulting in the spectrum shown in Figure 9. For every event, the ToF of the detected neutron can be determined and thus, knowledge of the source-to-detector distances facilitates the calculation of the neutron energy. The setup is self-calibrating thanks to physical events where two gamma-rays are emitted simultaneously by the source. These events both travel the well-known distances involved at the speed of light, and result in a gamma-flash in the ToF spectrum. The gamma-flash provides a reference point from which the instant of the double gamma-ray emission may be determined. In general, the source and YAP detectors are placed inside the central chamber of the Aquarium, while the neutron detector is placed at one of the beam ports.

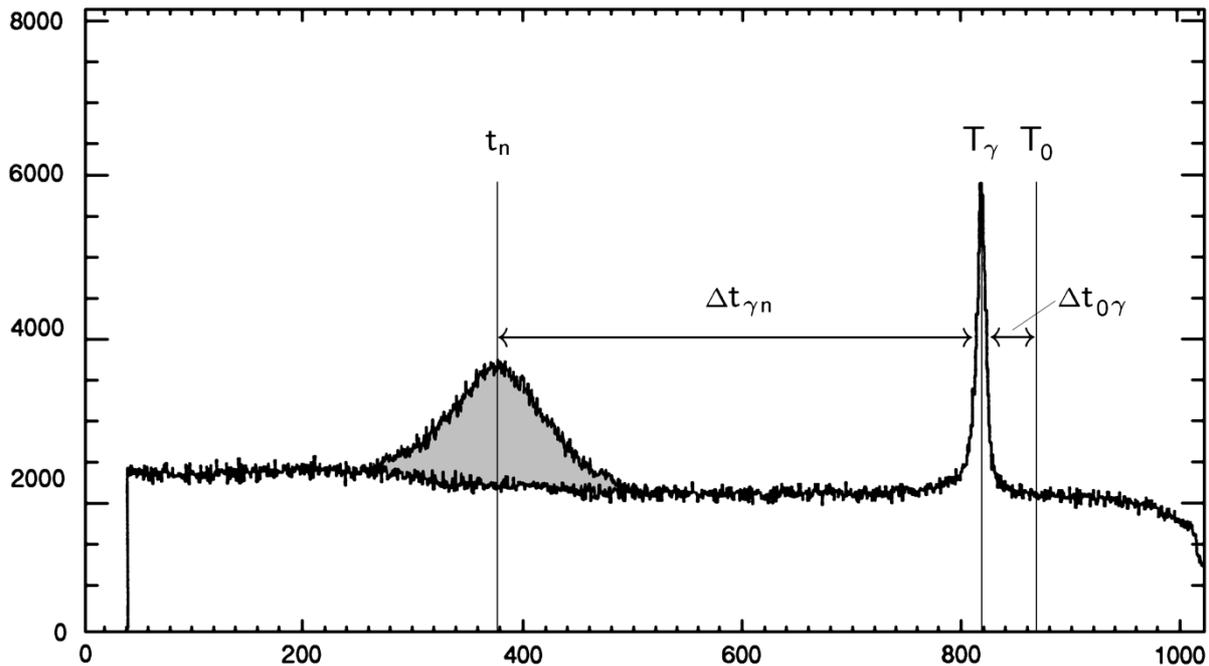

*Figure 9: TDC spectrum of tagged fast neutrons (TDC channels vs number of counts). An entry is the time between the start signal from the neutron detector and the stop signal from the gamma-ray detector. The gamma-flash is at $T_\gamma$, while the neutron distribution is at $t_n$. The indicated $T_0$-position is the reconstructed time of the actual decay. Figure from [19].*

With these techniques, the STF is presently capable of measuring the response of detectors to fast neutrons, fast-neutron detection efficiency and the neutron and gamma-ray attenuation properties of shielding materials (tagged neutron energy between 1 and 6 MeV). Note that a major upgrade of the Aquarium and the electronics available at the facility have recently been funded via the Lund University Natural Science Faculty. Together with a corresponding upgrade of the data-acquisition systems, we anticipate a first attempt to tag neutrons of energies in the thermal region (~ 25 meV) to commence very soon.

## 4. EXAMPLES FOR RECENT STUDIES PERFORMED AT THE STF

Highlights of recent results obtained by various user groups of the STF are presented below.

### 4.1. CHARACTERIZATION OF SOURCES

All of our neutron sources have been systematically characterized in-situ (see Figure 10), providing important validation benchmarks for our experimental infrastructure [20, 21]. This program of systematic characterization was deemed important since the sources were to be provided to a user community. Requiring each group in this user community to individually study the sources they were provided was felt to be unreasonably inefficient. The tagging technique has been extended to a Cf-252 fission fragment source [22]. For this, the single-sided Cf-252 source was positioned within a gaseous He-4 scintillator detector in which light and heavy fission fragments corresponding to neutron emission were detected as tags. Since the emission spectrum for Cf-252 is exceptionally well-known, an excellent benchmark exists from which it is anticipated the neutron-detection efficiency of a detector can be unfolded.

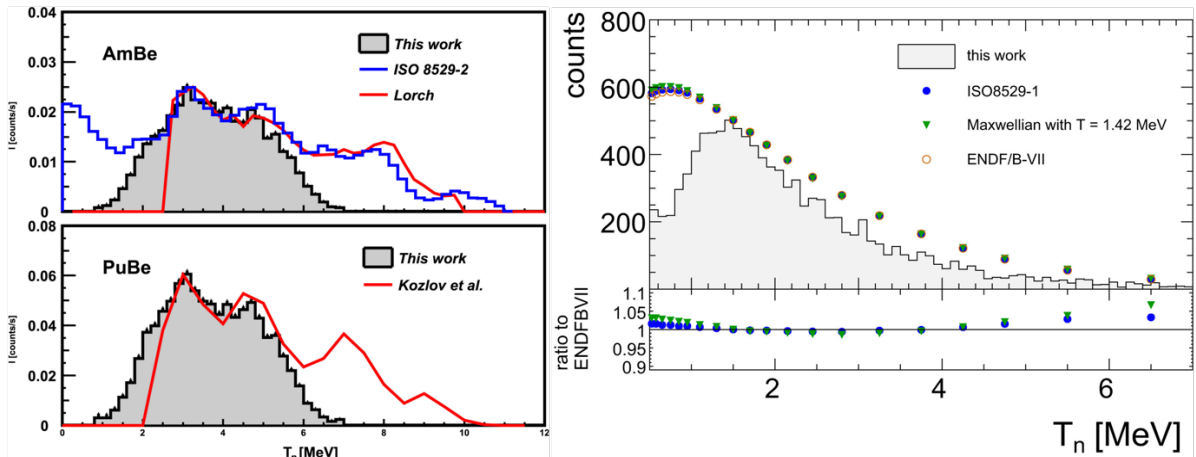

Figure 10: Spectra of sources measured at the STF. On the left, the measured spectra of the AmBe (top) and PuBe (bottom) sources (plots from [21]) and on the right, the measured spectrum of the Cf source (plot from [22]).

## 4.2. CHARACTERIZATION OF DETECTORS

The STF is ideal for the development of prototypes in preparation for tests at nuclear reactors or on real spallation instruments. It is anticipated that a very large subset of ESS detector prototypes will see the neutrons of the STF at some stage of their development (as the case for the Multi-Grid [23], to cite one). Using either direct irradiation or the tagging technique, simple functionality tests of detector prototypes may be performed. Further, the sensitivity of a prototype to fast-neutron or gamma-ray backgrounds may be investigated. Examples of the latter are recent studies performed on commercially available beam monitors [24] or on the Multi-Blade [25] detector (see Figure 11). The Multi-Blade detector is being developed for reflectometry instruments at ESS.

A recently developed black-box (recall Figure 4), to be equipped with a servo stages on an optical table, has been used to characterize Multi-Anode PMTs for the SoNDe project [26]. The SoNDe project is focused on the development of pixilated, solid-state neutron detectors for ESS.

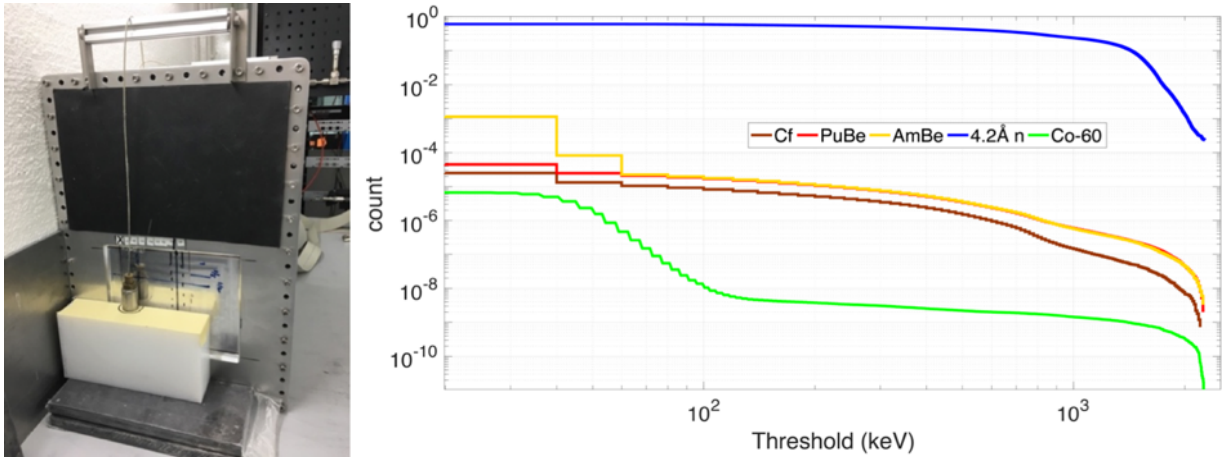

*Figure 11: Fast-neutron sensitivity measurement on the Multi-Blade detector: an example of untagged irradiation measure for detector characterization at STF. On the left, a photo of the detector irradiated by the PuBe source at STF and on the right, the cumulative number of counts as function of an energy threshold for different incoming radiations, normalised to the sensitivity of the detector to thermal neutrons (4.2Å). (Plot from [27]).*

### 4.3. CHARACTERIZATION OF SHIELDING

The potential of the STF is not limited to the development of new neutron instrumentation. Using its tagged-neutron beams, the behavior of materials under neutron or gamma-ray irradiation may be studied in detail [28]. This is a research domain traditionally addressed at reactors or spallation sources which has recently been invigorated with the promise of proton therapy in treating cancer. We investigated radiation attenuation in steel, copper, Polyethylene (PE) and both regular and PE/$B_4C$-enriched concrete samples using the Aquarium (see Figure 12). Measured transmission spectra of tagged neutrons were compared to simulation with a very high level of agreement (see Figure 13).

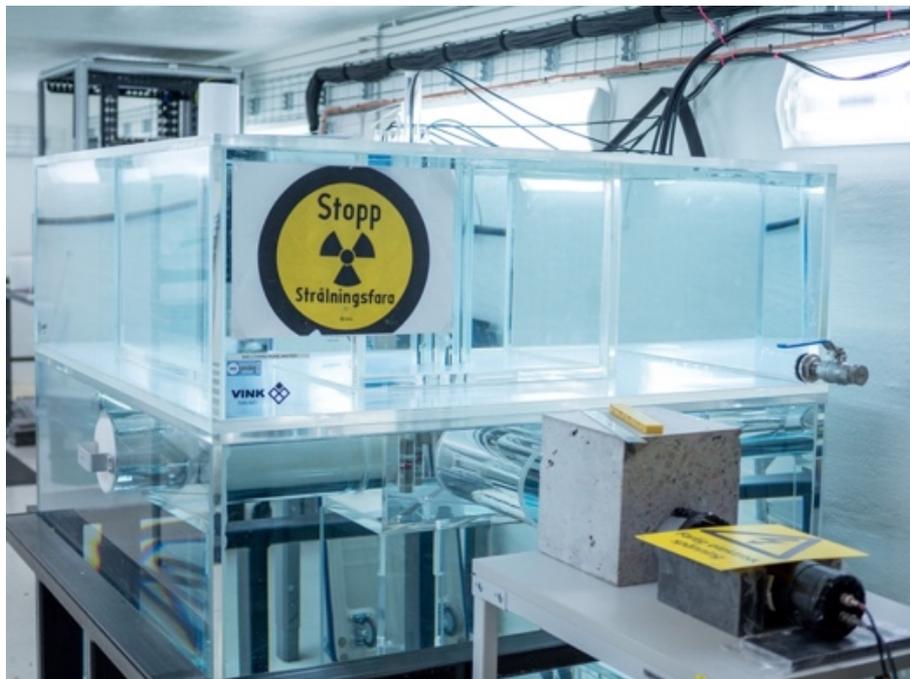

*Figure 12: Radiation attenuation in shielding material measurements. Transmission measurements were performed.*

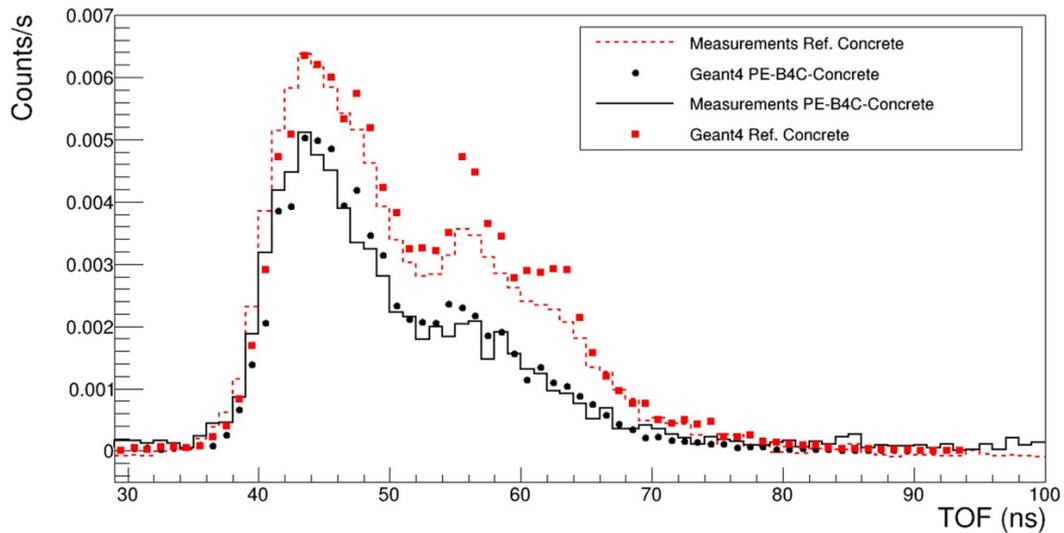

*Figure 13: A comparison of measurements and simulations for reference concrete and PE-$B_4$C-concrete. (Plot from [28]).*

## 4.4. EDUCATION

As part of Lund University infrastructure, the STF has been used extensively for student training at all levels. Table 1 shows a list of thesis work performed to date.

| Name of Student | Type of thesis | Institution | Date | Title |
|---|---|---|---|---|
| Amanda Jalgén | Master | LTH | September 12, 2017 | Initial Characterizations of a Pixelated Thermal-Neutron Detector |
| Laura Boyd | Summer | University of Glasgow | September 8, 2017 | Initial Testing of the Response of a Pixelated Thermal-Neutron Detector |
| Mohamad Akkawi | Summer | University of Toronto | June 13, 2017 | Photon Detection Using Cerium Bromide Scintillation Crystals |
| Nicholai Mauritzson | Master | LU | June 9, 2017 | Design, Construction and Characterization of a Portable Fast-Neutron Detector |
| Henrik Söderhielm | Under-graduate | LU | February 3, 2017 | Two-Dimensional Radiation Field Map of a Be-based Source |
| Julius Scherzinger | PhD | LU | December 16, 2016 | Neutron Irradiation Techniques |
| Emil Rofors | Master | LTH | March 14, 2016 | Fast Photoneutron Production |
| Sharareh Koufigar | Under-graduate | LU | October 21, 2015 | The Radiological Footprint of a Be-based Source |
| Julius Scherzinger | Licentitate | LU | March 20, 2015 | A Source-Based Testbed for Fast-Neutron Irradiation |

*Table 1: STF theses. (for a complete description, please visit http://www.nuclear.lu.se/english/research/neutronfysik/).*

## 5. SUMMARY

Located at the Department of Physics of Lund University, the Source-Testing Facility has been designed for advanced detector and material diagnostics. Being a user facility, the STF offers a complete set of infrastructure, including gamma-ray and neutron sources, shielding, detectors, computers and IT. Furthermore, acquisition, analysis, and simulation software, and support are offered by the SONNIG group, who operate the facility. Until now, the STF has been used almost exclusively for the development of He-3 free neutron detectors and the study of advanced neutron shielding. The STF offers the potential of a low cost, low barrier to entry, and low flux neutron source, that has potential applications beyond those utilised presently. The hands-on training of the next generation of neutron scientists is a high priority. If you are in need of neutrons, contact us. We are happy to provide access to the STF and support your measurements with consultations and hands-on support during beam-time.

- http://www.nuclear.lu.se/english/research/neutronfysik/
- https://europeanspallationsource.se/workshops-facilities#source-testing-facility
- stf@nuclear.lu.se
- Source Testing Facility c/o  
  Lund University  
  Department of Physics  
  Division of Nuclear Physics  
  P.O Box 118  
  SE-221 00 Lund  
  Sweden

**ACKNOWLEDGEMENTS**

This work has been funded by:

- the BrightnESS project, Work Package (WP) 4.4. BrightnESS is funded by the European Union Framework Program for Research and Innovation Horizon 2020, under grant agreement 676548.
- The Faculty of Science at Lund University (Grant for Infrastructure 2017, V2016/1949).
- The UK Science and Technology Facilities Council (Grant nos. STFC 57071/1 and STFC 50727/1).

**REFERENCES**

[1] SHEA D.A. and MORGAN D., "The Helium-3 shortage: supply, demand, and options for congress", Technical Report R41419, Congressional Research Service, (2010).

[2] KOUZES RT., "The 3He Supply Problem." PNNL-18388 Pacific Northwest National Laboratory, Richland, WA, (2009).

[3] LINDROOS M. et al., "The European Spallation Source", In Nuclear Instruments and Methods in Physics Research Section B, Volume 269, Issue 24 (2011), Pages 3258-3260, ISSN 0168-583X, https://doi.org/10.1016/j.nimb.2011.04.012. (http://www.sciencedirect.com/science/article/pii/S0168583X11003636 )

[4] PEGGS S. et al., "European Spallation Source Technical Design Report", ESS-2013-001

[5] HALL-WILTON R. et al, "Detectors for the European spallation source", IEEE Nuclear Science Symposium Conference Record, 4283-4289, (2012). doi:10.1109/NSSMIC.2012.6551977 url: http://ieeexplore.ieee.org/document/6551977/

[6] KIRSTEIN O. et al., "Neutron Position Sensitive Detectors for the ESS", (2014), PoS (Vertex2014) 029 arXiv:1411.6194

[7] CHERKASHYNA N. et al., "High energy particle background at neutron spallation sources and possible solutions", Journal of Physics: Conference Series 528, (2014), doi:10.1088/1742-6596/528/1/012013.

[8] http://www.nuclear.lu.se/english/

[9] https://europeanspallationsource.se/workshops-facilities#source-testing-facility

[10] https://confluence.esss.lu.se/display/DG/Detector+Group

[11] http://www.nuclear.lu.se/english/research/neutronfysik/

[12] BRUN R. and RADEMAKERS F., "ROOT - An Object Oriented Data Analysis Framework", Nucl. Instr. Meth. A, (1996) url: http://root.cern.ch/

[13] AGOSTINELLI S. et al, "Geant4 - a simulation toolkit", Nucl. Instr. Meth. A, (2003) url: http://www.sciencedirect.com/science/article/pii/S0168900203013688

[14] KITTELMANN T. et al, "Geant4 Based Simulations for Novel Neutron Detector Development", 20th International Conference on Computing in High Energy and Nuclear Physics (CHEP), (2013) doi:10.1088/1742-6596/513/2/022017; arXiv:1311.1009v1.

[15] KITTELMANN T. and BOIN M., "Polycrystalline neutron scattering for Geant4: NXSG4", Computer Physics Communications 189, 114-118; (2015). doi:10.1016/j.cpc.2014.11.009.

[16] CAI XX and KITTELMANN T., "NCrystal : A library for thermal neutron transport in crystals", http://mctools.github.io/ncrystal/ v0.9.1, https://doi.org/10.5281/zenodo.855292

[17] SCHERZINGER J., "Neutron Irradiation Techniques", Lund University, Faculty of Science, Department of Physics, (2016)  url = http://portal.research.lu.se/portal/en/publications/neutron-irradiation-techniques(d4c447d5-ee52-49a9-ad55-4669aed57e32).html

[18] SCHERZINGER J. et al, "The light-yield response of a NE-213 liquid-scintillator detector measured using 2–6 MeV tagged neutrons", (2016) url = http://www.sciencedirect.com/science/article/pii/S0168900216310361

[19] NILSSON B., "High-resolution Measurement of the $^4$He ($\gamma$, n) Reaction in the Giant Resonance Region", Lund University, Faculty of Science, Department of Physics, (2003) url = http://portal.research.lu.se/portal/en/publications/highresolution-measurement-of-the-4hegn-reaction-in-the-giant-resonance-region(f9873930-d37b-4e4b-b6d0-1acba29ee46e).html

[20] SCHERZINGER J. et al, "Tagging fast neutrons from an $^{241}$Am/$^9$Be source", (2015) url = http://www.sciencedirect.com/science/article/pii/S0969804315000044

[21] SCHERZINGER J. et al, "A comparison of untagged gamma-ray and tagged-neutron yields from $^{241}$AmBe and $^{238}$PuBe sources", (2017) url = http://www.sciencedirect.com/science/article/pii/S0969804316309861

[22] SCHERZINGER J. et al, "Tagging fast neutrons from a $^{252}$Cf fission-fragment source", (2017) url = http://www.sciencedirect.com/science/article/pii/S0969804316310521

[23] KHAPLANOV A. et al, "10B multi-grid proportional gas counters for large area thermal neutron detectors", Nucl. Instr. Meth. A, Vol 720, Pag 116-121, (2013), ISSN 0168-9002, https://doi.org/10.1016/j.nima.2012.12.021.

[24] ISSA F. et al, "Characterization of Thermal Neutron Beam Monitors", Phys. Rev. Accel. Beams 20, 092801 (2017) url = https://arxiv.org/abs/1702.01037

[25] PISCITELLI F. et al., "The Multi-Blade Boron-10-based neutron detector for high intensity neutron reflectometry at ESS", JINST, (2017) url = http://iopscience.iop.org/article/10.1088/1748-0221/12/03/P03013/meta;jsessionid=ADCECAD7D2492490212685E84DC81C65.c4.iopscience.cld.iop.org#references

[26] JAKSCH S. et al, "Cumulative Reports of the SoNDe Project July 2017", arXiv preprint arXiv:1707.08679 (2017), url = http://www.fz-juelich.de/jcns/jcns-2/EN/Forschung/Instruments-for-ESS/SoNDe-Projekt/_node.html

[27] MESSI F. et al, "Gamma- and Fast Neutron- Sensitivity of 10B-based Neutron Detectors at ESS", 2017 IEEE Nuclear Science Symposium and Medical Imaging Conference (2017 NSS/MIC)

[28] DIJULIO D. et al, "A polyethylene-$B_4C$ based concrete for enhanced neutron shielding at neutron research facilities", (2017) url = http://www.sciencedirect.com/science/article/pii/S0168900217304151